\documentclass[preprint2]{aastex}
\usepackage{graphicx}
\usepackage{amsmath,amssymb}
\usepackage{epstopdf}

\usepackage{natbib}
\newcommand{\vect}[1]{\mathbf{#1}}
\usepackage{ulem}

\DeclareMathOperator{\sech}{{\rm sech}}

\begin{document}

\title{The three-dimensional evolution of ion-scale current sheets: tearing and drift-kink instabilities in the presence of proton temperature anisotropy}
\author{P. W. Gingell, D. Burgess}
\affil{Queen Mary, University of London}
\affil{Mile End Road, London, E4 1NS, UK}
\email{p.w.gingell@qmul.ac.uk}
\and
\author{L. Matteini}
\affil{Imperial College London}
\affil{London, SW7 2AZ, UK}

\begin{abstract}

We present the first three-dimensional hybrid simulations of the evolution of ion-scale current sheets, with an investigation of the role of temperature anisotropy and associated kinetic instabilities on the growth of the tearing instability and particle heating. We confirm the ability of the ion cyclotron and firehose instabilities to enhance or suppress reconnection, respectively. The simulations demonstrate the emergence of persistent three-dimensional structures, including patchy reconnection sites and the fast growth of a narrow-band drift-kink instability, which suppresses reconnection for thin current sheets with weak guide fields. Potential observational signatures of the three-dimensional evolution of solar wind current sheets are also discussed. We conclude that kinetic instabilities, arising from non-Maxwellian ion populations, are significant to the evolution of three-dimensional current sheets, and two-dimensional studies of heating rates by reconnection may therefore over-estimate the ability of thin, ion-scale current sheets to heat the solar wind by reconnection.

\end{abstract}


\section{Introduction}

The solar wind is one of the most accessible laboratories for the study of kinetic plasma turbulence and associated plasma instabilities. Recent progress owed to solar and terrestrial orbiting satellites has led to valuable gains in the understanding of physical processes in the solar wind, however a full solution to the solar wind heating problem remains elusive. The combination of observational data with analytic and numeric analysis of turbulent systems and solar wind structures have led to significant strides in our understanding of the role of wave-particle interactions and reconnection in solar wind heating \citep{turb_living_review}.

Measurements of ion temperatures in the solar wind and magnetospheric plasma environments reveal the signatures of kinetic plasma instabilities driven by the observed temperature anisotropies \citep{bale2009,wicks2013}. These processes generate unstable fluctuations which can interact with particle populations, scattering them towards the more isotropic, marginal stability state. It has been demonstrated that these microphysical instabilities strongly affect the macroscopic evolution of the host plasma, in this case the solar wind \citep{matteini2012rev,matteini2013jgr}. Similar processes are also expected to be active in astrophysical plasmas, such as galaxy clusters (e.g. \citealt{schekochihin2005}) and accretion disks \citep{sharma2007,riquelme2012,kunz2014}.

The solar wind has also been observed to host a population of magnetic discontinuities \citep{erdos2008}, mostly of rotational geometry with small but significant fraction observed to be tangential \citep{chou2012}. These current sheet structures are potential sites of magnetic reconnection \citep{servidio2011}, and appear to be associated with local temperature enhancements \citep{osman2012b}, suggesting that they may play a role in plasma heating. However, the fraction of observations of discontinuities associated with identified reconnection events is small \citep{gosling2007}, and their contribution to solar wind heating is therefore not well understood. A lack of correlation between strong current sheets and preferential plasma heating \citep{borovksy2011} suggests that properties other than current sheet thickness are important in determining their stability and evolution, and therefore any associated heating. However, recently observed correlations between proton temperature anisotropies and current sheet events \citep{osman2012a} invites a study of coupling between kinetic instabilities and solar wind structures such as current sheets. We note that differences between these correlations may arise due to the authors' definitions of what constitutes an observation of a current sheet. Numerical simulations of tearing have been able to demonstrate that reconnection can influence local particle distributions, including generation of temperature anisotropies \citep{aunai2011,servidio2012} and particle acceleration \citep{drake2010}.

The role of temperature anisotropy in the tearing instability of current sheets has seen limited investigation. Early studies of the linear theory have predicted that the reconnection rate may be enhanced in the presence of temperature anisotropy $T_\perp>T_\parallel$, with the fastest growing mode shifted to smaller wavelengths, and suppressed for $T_\parallel > T_\perp$ \citep{chen1984}. This has been confirmed by numerical models for hybrid \citep{ambrosiano1986} and fluid \citep{shi1987} regimes. Most recently, \citep{matteini2013} has demonstrated this principle for ion-scale current sheets in two-dimensions using a fully self-consistent hybrid model. It was shown that the growth of ion cyclotron waves (for $T_\perp>T_\parallel$) and the firehose instability (for $T_\parallel > T_\perp$) in the background perturbs the current sheet in such a way as to pinch or kink the current sheet, respectively enhancing or suppressing the reconnection process. 

In this Paper, we present the first three-dimensional hybrid simulations exploring the dependence of the evolution of the ion-scale tearing instability on proton temperature anisotropies. Having confirmed the change in reconnection rate for background temperature anisotropies found in two-dimensional simulations, we demonstrate the same relationship between the growth of kinetic instabilities in the background plasma and the tearing of three-dimensional current sheets. Furthermore, we demonstrate the evolution of three-dimensional structures, including the fast growth of a drift-kink instability and the late-time appearance of patchy reconnection sites.

\section{Simulations}

We investigate the evolution of ion scale current sheets by means of simulations utilizing a hybrid model, which combines a kinetic, particle-in-cell treatment of ion species with a charge-neutralising, massless and adiabatic electron fluid \citep{matthews1994}. Maxwell's equations are solved under the low-frequency Darwin limit, neglecting collisions and with zero resistivity. The principal simulations presented in this paper are three-dimensional in configuration space, though we also present 2.5-D simulations (i.e. a 2-D grid with 3 vector components for fields and velocities, e.g. $B_{x,y,z}(x,y,t)$) for comparison where appropriate.

Simulations are initialised in Harris equilibrium \citep{harris1962} in the $x$-direction, with current-carrying direction $z$:

\begin{eqnarray}
\label{eqn:bxcond} B_x &=& 0\\
B_y(x) &=& B_0\sin\left(\theta/2\right)\tanh\left(x/L\right)\\
B_z &=& B_0\cos\left(\theta/2\right)\\
\label{eqn:ncond} n(x) &=& n_0 +n_{\rm cs}\sech^2\left(x/L\right),
\end{eqnarray}

\noindent where $B_0$ is the background magnetic field strength, $n_0$ is the background proton number density, $n_{\rm cs}$ is the peak number density of the current sheet population, $L$ is the current sheet width, and $\theta$ is the angle between background magnetic fields either side of the current sheet. The case $\theta = \pi$ represents an anti-parallel geometry with no guide field.

A requirement of zero electric field (i.e. pressure balance) across the current sheet introduces the condition

\begin{equation}
P_{\rm \perp,e}+P_{\rm \perp,p}+\frac{B^2}{8\pi} = {\rm Const.},
\end{equation}
\noindent which, with Equations \ref{eqn:bxcond}-\ref{eqn:ncond} and perpendicular gas pressures $P_{\rm \perp,s} = n_{\rm s}kT_{\rm \perp,s}$, can be simplified to
\begin{equation}
\beta^{\rm cs}_{\rm \perp,e} + \beta^{\rm cs}_{\rm \perp,p} = 1,
\label{eqn:beta_cond}
\end{equation}

\noindent where $\beta^{\rm cs}_{\rm \perp,s} = 2\mu_0 n_{\rm cs}kT^{\rm cs}_{\perp, s}/B_0^2$. For a Harris equilibrium for which the current is carried solely by the drifting ion population, the drift speed of this population is given by:

\begin{equation}
v_{{\rm p},0} = -\frac{w_\perp\rho_{\rm p}}{L},
\label{eqn:ion_current}
\end{equation}

\noindent where $w_\perp$ is the perpendicular thermal velocity of the drifting ion population, and $\rho_{\rm p}$ is the proton gyro-radius.

Three-dimensional simulations presented in this paper are extensions of the geometry in \citet{matteini2013} in the $z$-direction, featuring a $(N_x, N_y, N_z) = (256, 128, 256)$ grid, with resolution $\Delta x,z = 0.5d_i$ and $\Delta y = d_i$, where $d_i = v_A/\Omega$ is the ion inertial length. Spatial and temporal scales are given in units of the ion inertial length $d_i$ and the inverse ion gyrofrequency $t_\Omega = \Omega_i^{-1} = (qB_0/m)^{-1}$ respectively. Two current sheets, with current carrying directions $\pm z$, are included in the initial conditions, as necessitated by the use of triply-periodic boundary conditions. Two-dimensional simulations are flattened in the $x$ or $z$-directions (i.e., $N_x = 1$ or $N_z = 1$) as necessary to isolate two-dimensional processes in those planes. We make use of at least 100 computational particles per cell to sufficiently sample the distribution function in the tail regions of velocity space.

The perpendicular plasma beta is given by $\beta_{\rm \perp} = 0.5$ for all simulations and for both background and drifting ion populations, with variation in $\beta_{\rm \parallel}$ defining a background temperature anisotropy $A_0 = T_\perp/T_\parallel$. The electron fluid is isotropic with $\beta_e = 0.5$. Fields and moments are normalised such that $n_0 = 1$ and $B_0 = 1$, with $n_{\rm cs} = 1$ for the current sheet population. Note that these parameters present a slightly lower plasma beta than the previous two-dimensional study of the geometry by \citealt{matteini2013}.
The parameter space sampled by the simulations presented in this paper, including variations in temperature anisotropies and field angle, is summarised in Table \ref{tab:allsims}.

\begin{table*}
\caption{\label{tab:allsims}List of simulation parameters.}
\begin{tabular}{ccc}
Anisotropy, $A_0 = T_{\perp}/T_{\parallel}$& Field angle, $\theta/\pi$ & Dimensionality\\
\tableline
1  & 1 & 3-D, 2-D(x,y), 2-D(x,z)\\
2  & 1 & 3-D, 2-D(x,y), 2-D(x,z)\\
0.5 & 1 & 3-D, 2-D(x,y), 2-D(x,z)\\
1 & 0.9 & 3-D, 2-D(x,z) \\
1  & 0.8 & 3-D, 2-D(x,z)\\
1  & 0.7 & 3-D, 2-D(x,z)\\
\end{tabular}
\end{table*}

\section{Results}

\begin{figure*}
\includegraphics[width=\linewidth,clip,trim = 20 0 20 0]{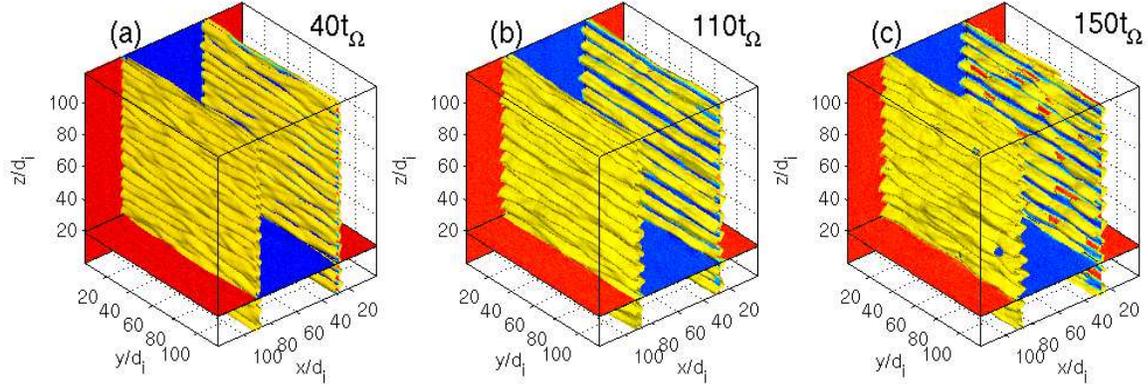}
\caption{\label{fig:evo3d_isosurfs_dens} 
Isosurfaces of current sheet proton number density $n=0.2n_0$ at times $t=40,110,150t_\Omega$, for a simulation initialised with $A_0=1, \theta = \pi$. The tangential magnetic field $B_y$ is also shown in slice planes at $x/d_i= 30$, $y/d_i = 0$ and $z/d_i = 20$.
These snapshots demonstrate early kink of the current sheet, and appearance of patchy reconnection sites at late times with depleted current sheet density.}
\end{figure*}

\begin{figure}
\includegraphics[width=\linewidth,clip,trim = 0 20 0 25]{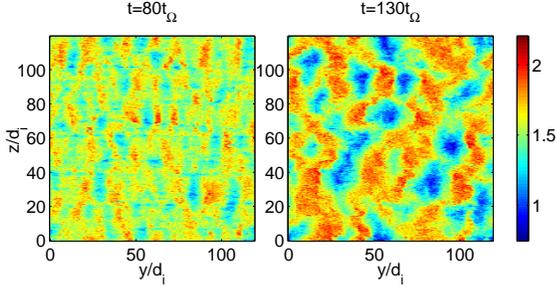}
\caption{\label{fig:flux_tubes} 
Proton number density on the kinked, 3-D $B_y=0$ surface, visualised as a planar projection onto the $y-z$ plane, for a 3-D simulation with $A_0=2, \theta = \pi$. At $t=80t_\Omega$, we see the beginning growth of flux tubes extended in the $z$-direction, with reconnecting x-lines following the low-density regions. At $t=130t_\Omega$, these flux tubes are seen to have merged into larger structures with a kink in the field-aligned $y$-direction, indicating the break up of x-lines into patchy reconnection sites.}
\end{figure}

\begin{figure*}
\includegraphics[width=\linewidth,clip,trim = 20 0 20 0]{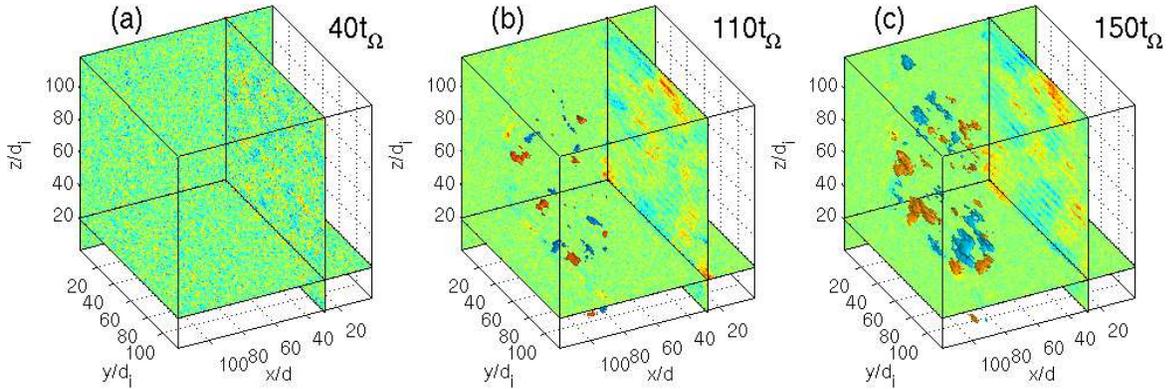}
\caption{\label{fig:evo3d_isosurfs_vy} Current density $J_y$ for $A_0=1, \theta = \pi$ at times $t=40,110,150t_\Omega$, with a slice through the null B-field plane of one current sheet, and isosurfaces $J_y = \pm 0.1 en_0v_A$ for the other. Note the late-time appearance of reconnection exhausts at patchy reconnection sites.}
\end{figure*}

The three dimensional evolution of the current sheets for a simulation with parameters $(A_0,\theta) = (1,\pi)$ is presented in Figures \ref{fig:evo3d_isosurfs_dens} and \ref{fig:evo3d_isosurfs_vy}. These figure demonstrates several key features of the evolution, common to the majority of cases described in Table \ref{tab:allsims}, which can be summarised as follows:

\begin{enumerate}
\item The number density isosurfaces in Figure \ref{fig:evo3d_isosurfs_dens}(a) show the growth of a narrow-band wave with $\vect{k}\parallel\vect{J}$ at early times ($t<40t_\Omega$), identified as a drift-kink instability and described further in Section \ref{sec:driftkink}. The persistence of this feature in Figures \ref{fig:evo3d_isosurfs_dens}(b) and \ref{fig:evo3d_isosurfs_dens}(c) suggest this instability dominates the evolution under the chosen parameters.
\item Sites of depleted current sheet number density in Figure \ref{fig:evo3d_isosurfs_dens}(c), e.g. $(x,y,z) = (30,80,60)$, are suggestive of the generation of current-aligned flux tubes by reconnection, indicating the slow growth of a tearing instability at late times. 
\item Number density on the kinked, 3-D  $B_y=0$ surface, visualised as a planar projection onto the $y-z$ plane, is shown for a simulation with $A_0 = 2$ in Figure \ref{fig:flux_tubes}. These projections demonstrate the generation of quasi-2-D, current-aligned flux tubes and associated x-lines more clearly than in Figure \ref{fig:evo3d_isosurfs_dens}(c). At $t=80t_\Omega$, these flux tubes are seen to extend approximately $60d_i$ in the $z$-direction, and the evolution is therefore similar to what would be expected for the linear phase of the two-dimensional tearing instability over that length scale \citep{matteini2013}. At $t=130t_\Omega$, the flux tubes are seen to have merged and to kink in the field-aligned $y$-direction, indicating the development of patchy, three-dimensional reconnection sites. The growth of the tearing instability is discussed further in Section \ref{sec:tearing}.
\item At late times, in Figures \ref{fig:evo3d_isosurfs_vy}(b) and \ref{fig:evo3d_isosurfs_vy}(c), we see the growth of field-aligned velocity enhancements visible in slice planes and isosurfaces of $J_y$ within the current sheet. These enhancements occur at the same positions as the number density reductions in Figure \ref{fig:evo3d_isosurfs_dens}(c) and Figure \ref{fig:flux_tubes}, providing evidence of particle acceleration by reconnection due to the tearing instability. The short extent of these structures in the $z$-direction is consistent with the growth of patchy reconnection sites, rather than extended x-lines. Particle heating by reconnection is discussed further in Section \ref{sec:heating}.

\end{enumerate}

\subsection{\label{sec:tearing}Tearing Instability}

\begin{figure}
\includegraphics[width=\linewidth,clip,trim = 000 0 0 000]{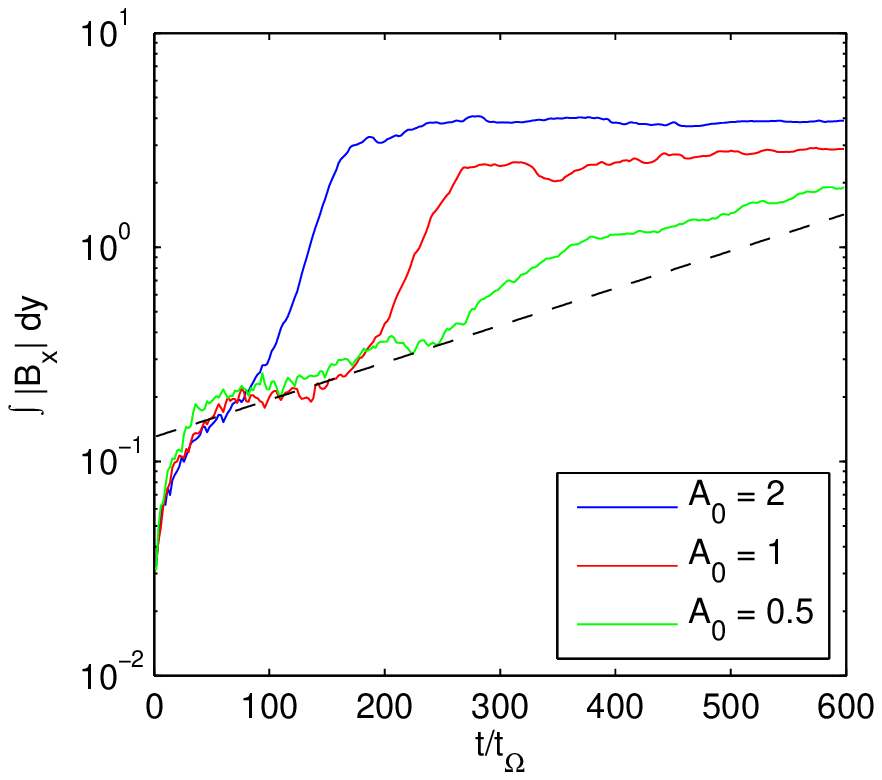}
\caption{\label{fig:2d_recon_rate} Temporal evolution of the reconnected flux for two-dimensional simulations with varying background temperature anisotropy $A_0$. Higher perpendicular temperature $A_0 > 1$ results in faster reconnection, while higher parallel temperature $A_0 < 1$ results in slower reconnection.}
\end{figure}

An estimate of the reconnection rate associated with the tearing instability is determined by calculating the reconnected flux, i.e. integrating the magnitude of the component of the magnetic field perpendicular to the current sheet, $\left|B_x\right|$, over the planes corresponding to the initial null field planes $B_y = 0$ at $x = \tfrac{1}{4}N_x\Delta x, \tfrac{3}{4}N_x\Delta x$ . Using a set of two-dimensional simulations of the $x-y$ plane, we first confirm the relationship discussed by \citealt{chen1984} and reproduced in \citealt{matteini2013}: that a faster tearing instability, and therefore faster reconnection rate, is seen for $A_0>1$, and a weaker instability is seen for $A_0<1$ (Figure \ref{fig:2d_recon_rate}).

\begin{figure}
\centering
\includegraphics[width=\linewidth,clip,trim = 25 0 25 0]{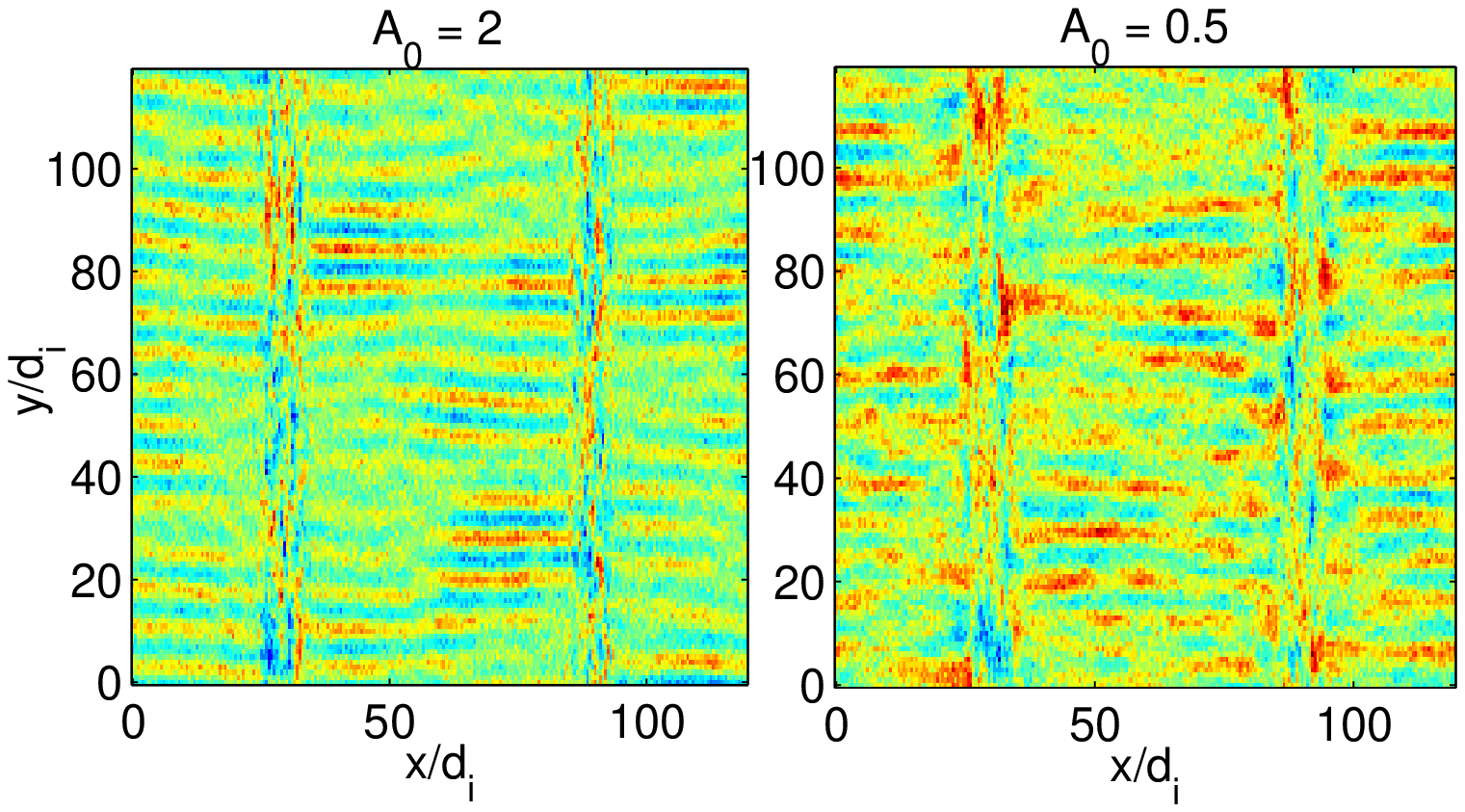}
\caption{{\label{fig:back_inst}} Background fluctuations in the out-of-plane magnetic field component $B_z$ grow as a result of the ion cyclotron instability (shown left, for $A_0=2$ at $t=100t_\Omega$ in the plane $z=0$ in a 3-D simulation) and the firehose instability (shown right, for $A_0=0.5$ at $t=100t_\Omega$ in the plane $z=0$ in a 3-D simulation). Correlated and anti-correlated fluctuations either side of the current sheet enhance or suppress tearing in the $A_0>1$ and $A_0<1$ cases respectively.}
\end{figure}

For $A_0>1$ in the 2-D and 3-D simulations, we observe at early times the growth of an ion-cyclotron mode in the background, which propagates in the local magnetic field direction. Forced perturbation of the current sheet by these modes, symmetric on either side of the current sheet, leads to efficient triggering of the tearing instability \citep{matteini2013}. Conversely, for $A_0<1$ and sufficiently high plasma beta, oscillatory modulation of magnetic field lines due to the growth of a firehose instability in the background plasma leads to kinking of the current sheet and suppression of reconnection. At late times, a new equilibrium is achieved, with a thicker current sheet, which can undergo tearing with reduced growth rate compared to an isotropic initial condition. The fluctuations associated with the ion cyclotron and firehose instabilities are shown in Figure \ref{fig:back_inst}.

\begin{figure}
\includegraphics[width=\linewidth,clip,trim = 000 0 0 000]{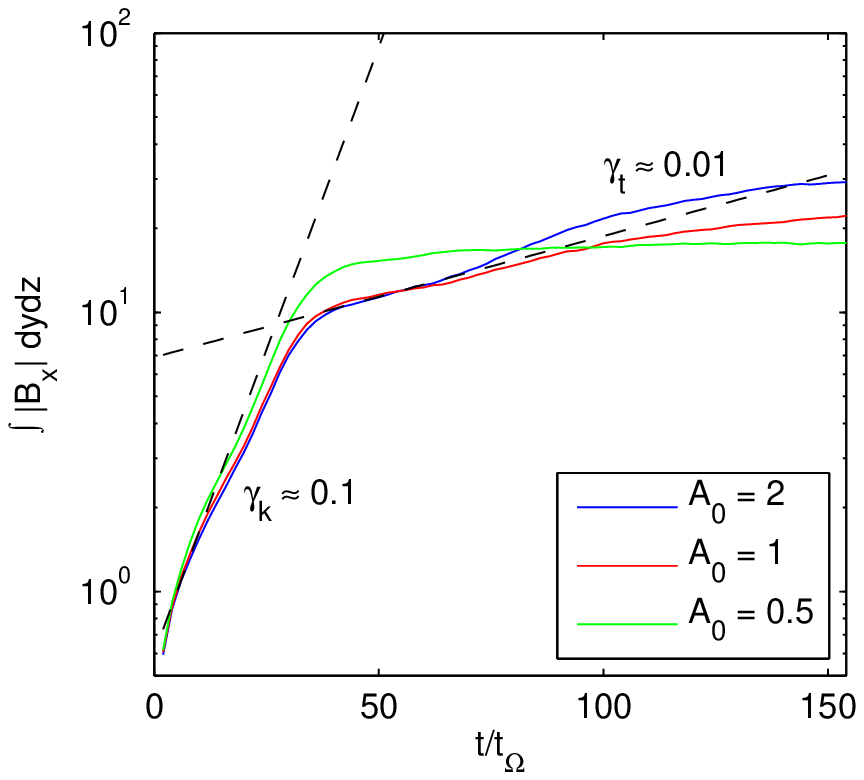}
\caption{\label{fig:3d_recon_rate} Temporal evolution of the reconnected flux for three-dimensional simulations with varying background temperature anisotropy $A_0$. After a fast period of growth of $|B_x|$ attributed to the drift-kink instability, reconnection is observed from $t>50t_\Omega$. As in the 2-D case, the reconnection rate is faster for $A_0>1$ and slower for $A_0<1$. The approximate linear growth rates are shown for the drift-kink ($\gamma_k$) and tearing ($\gamma_t$) instabilities.}
\end{figure}

\begin{figure}
\includegraphics[width=\linewidth,clip,trim = 0 10 10 9]{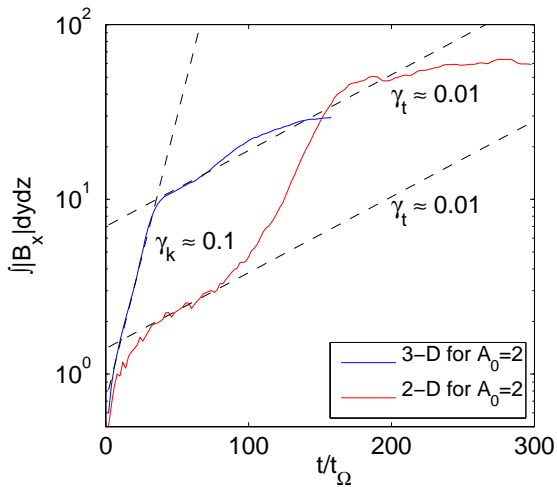}
\caption{\label{fig:2dvs3d_recon_rate} Comparison of the evolution of reconnected flux for 2-D and 3-D simulations with $A_0=2$. Note that the growth rate associated with the tearing phase of the evolution of the 3-D simulation is reduced compared to the 2-D case, and the non-linear phase of island merging is delayed. The approximate linear growth rates are shown for the tearing ($\gamma_t$) instability for both 2-D and 3-D cases, and also for the drift-kink ($\gamma_k$) instability for the 3-D case.}
\end{figure}

The same analysis applied to the full three-dimensional simulations gives the set of reconnection rates shown in Figure \ref{fig:3d_recon_rate}. As with the two-dimensional simulations, wave growth due to kinetic instabilities in the background plasma leads to a change in the growth rate of the tearing instability at intermediate times, and in the time at which the non-linear phase of the instability begins, during which magnetic islands merge. A comparison of the evolution of the reconnected flux between isotropic two- and three-dimensional cases, Figure \ref{fig:2dvs3d_recon_rate}, reveals i) for 3-D simulations, we observe an initial growth in the perpendicular magnetic flux which we attribute to the drift-kink instability described in Section \ref{sec:driftkink}, and ii) the non-linear phase of the tearing instability begins at a later time for the 3-D simulation. The latter point is a consequence of the strong limit on the growth of the magnetic islands in the plane perpendicular to the current, caused by the dynamics introduced with the extension to three dimensions.

\subsection{\label{sec:driftkink}Drift-Kink Instability}

\begin{figure}
\includegraphics[width=\linewidth]{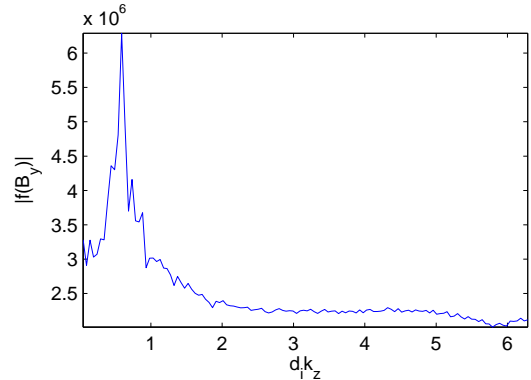}
\caption{\label{fig:dk_spectrum}Narrow-band spectrum in $k_z$ of the drift-kink instability for $A_0=1$, from the Fourier transform of the tangential magnetic field $B_y$.}
\end{figure}

The clearest evolutionary feature which develops with extension to three dimensions is the kink of the current sheets in the current-carrying direction at early times, before reconnection due to the tearing instability has begun. We attribute this evolution to the ion-ion drift-kink instability \citep{daughton1998}, a stream-type instability driven by the relative drift between the background ion species and the current carrying population at the magnetic discontinuity. Linear analysis of the drift-kink instability utilizing a Vlasov description of both electrons and ions at thin current sheets \citep{daughton1999} suggests a strong dependence of the growth rate of the instability on current sheet thickness, which is proportional to the ion current via Equation \ref{eqn:ion_current}. This suggests that this instability may only be important to the evolution of three-dimensional current sheets for discontinuities at ion-scales and below. For thicker, fluid-scale discontinuities, the growth rate of the drift-kink is small compared to the tearing mode for realistic mass ratios \citep{daughton1999}. However, we note that differences between our simulations and these studies may arise due to the neglect of electron-scale physics in the hybrid model. Although kink of fluid-scale current sheets has been observed with power in a broad spectrum of modes in $k_z$ (e.g. \citealt{landi2012}), we note that here the instability grows with a narrow-band or monochromatic spectrum, as shown in Figure \ref{fig:dk_spectrum}, consistent with other kinetic simulations such as \citealt{yin2008}.

\begin{figure}
\includegraphics[width=\linewidth]{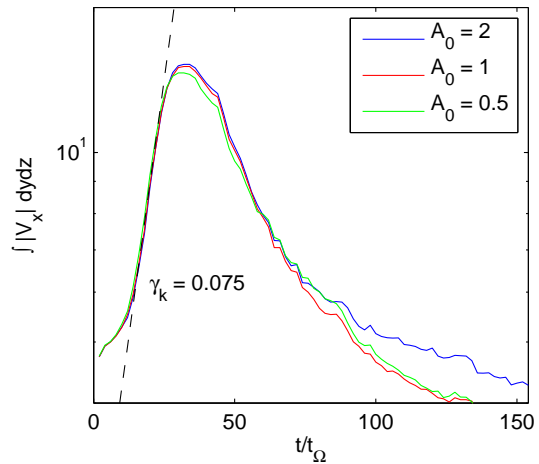}
\caption{\label{fig:dk_growthrate} Growth of the drift-kink instability for varying temperature anisotropy, from the perpendicular velocity flux $\int|v_x|dydz$, with linear growth rate $\gamma_k$.}
\end{figure}

The growth rates of the drift-kink instability compared to the tearing instability are shown for several simulations in Figure \ref{fig:dk_growthrate}. 
Unlike the tearing instability, we find that the drift-kink is unaffected by the background temperature anisotropy. 
For sufficiently high anisotropy $A_0$, it has been shown that the growth rate of the tearing instability is independent of sheet thickness, instead being forced by fluctuations generated by the growth of the ion cyclotron instability \citep{matteini2013}. Hence, for thicker current sheets with high background anisotropies, the tearing instability could initially dominate the evolution, before the drift-kink instability grows to an appreciable level.
In such a case, we expect to observe 3-D kink of initially 2-D flux tubes formed by reconnection, rather than kink of the initially planar current sheet.

\begin{figure}
\includegraphics[width=\linewidth]{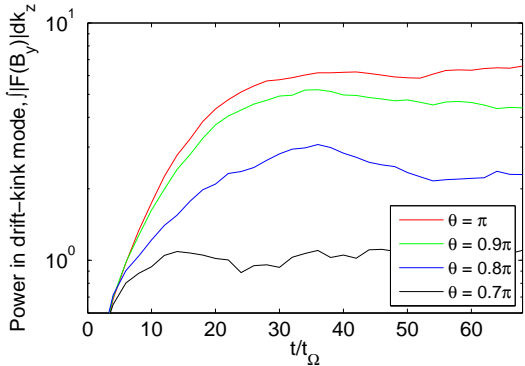}
\caption{\label{fig:dk_growth_guide} Dependence of the growth rate of the drift-kink instability on guide field angle, from the power in $k_z$ from fluctuations in $B_y$ in the initially null-field plane. Note the fast reduction in growth rate for small deviations from anti-parallel geometry.}
\end{figure}

We also note that the drift-kink instability is strongly suppressed by a guide field, as shown in Figure \ref{fig:dk_growth_guide}, with no growth seen for angles $\theta<0.7\pi$. Hence, observations of thin, kinked current sheets are likely to be rare in the solar wind, where turbulent fluctuations are expected to generate a population of current sheets with a broad range of guide fields \citep{borovksy2008,miao2011,zhdankin2012}.

Importantly, we note that drift-kink and tearing instabilities are not wholly separable: in the three-dimensional case we observe a reduction in the growth rate of the tearing instability after saturation of the drift-kink instability. This occurs due to the effective widening of the current sheet by the drift-kink instability, similar to the effect of the perpendicular kink caused by the firehose instability in the $A_0<1$ case.

\subsection{\label{sec:heating}Ion Heating}

\begin{figure}
\includegraphics[width=\linewidth]{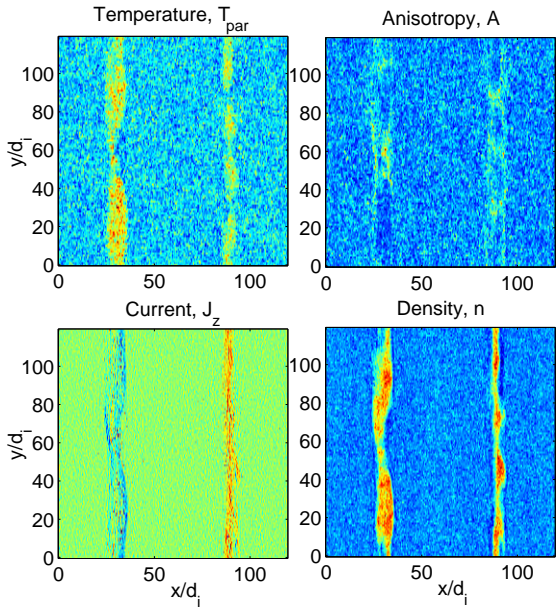}
\caption{\label{fig:tempmaps}$T_\parallel$ (left) and temperature anisotropy $T_\perp/T_\parallel$ (right) at $t=150t_\Omega$ for a slice of a 3-D simulation initialised with $A_0 = 1, \theta = \pi$. The current $J_z$ and total number density $n$ are also given to demonstrate the position of reconnection sites and flux tubes. Note that we see temperature enhancements in reconnection exhausts within flux tubes, and high temperature anisotropy at reconnection sites.}
\end{figure}

In order to assess the impact of three-dimensional reconnection, and other associated kinetic instabilities, on particle energisation and heating in solar wind ion-scale current sheets, we examine the temporal evolution of moments of the phase space distribution functions for ions in the current sheet's drifting population. Plots of temperature and temperature anisotropy variations across the current sheet are provided in Figure \ref{fig:tempmaps} for three-dimensional simulations. These plots demonstrate the appearance of temperature enhancements in the exhaust regions, and an increase in temperature anisotropy at the reconnection sites. Although no explicit temperature enhancements or depressions are visible resulting from the drift-kink instability, we note that temperature enhancements in reconnection exhausts appear later in the three-dimensional cases compared to the two-dimensional simulations, consistent with the reduction in the growth rate of the tearing instability from two to three dimensions caused by the drift-kink. 

\begin{figure}
\includegraphics[width=\linewidth]{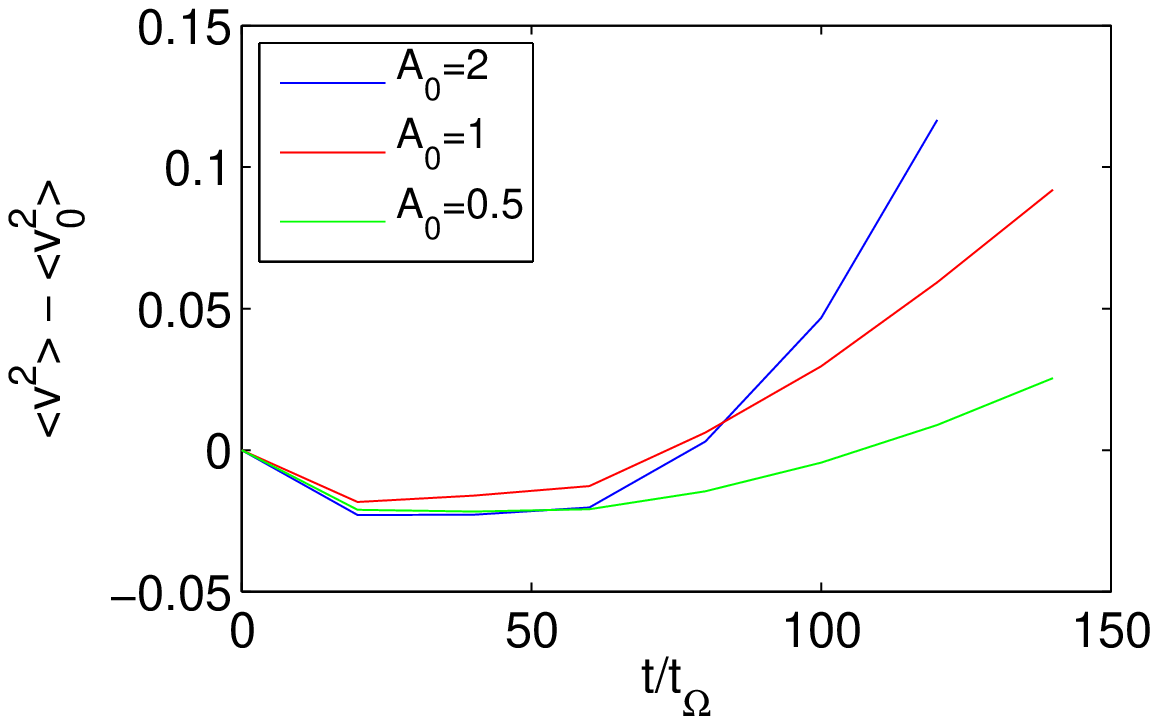}
\caption{\label{fig:part_energy} Change in the mean kinetic energy of ions within $10d_i$ of the initial current sheet planes, demonstrating faster energisation of ions in the case $A_0>1$ and slower energisation of ions in the case $A_0<1$, consistent with the measured growth of the tearing instability.}
\end{figure}

The mean kinetic energy of ions in the current sheet population is shown in Figure \ref{fig:part_energy}. As with the temperature enhancement maps, we note a significant reduction in ion heating in the three-dimensional cases for all variations in temperature anisotropy. As expected, the variation in rate of increase of particle energy matches the growth of the tearing instability as the temperature anisotropy is changed.

We are therefore able to conclude that reconnection at ion-scale current sheets may contribute to the increase of particle energies in the solar wind, visible as strong beams or temperature enhancements in the reconnection exhausts. However, these effects are strongest for current sheets with close to anti-parallel geometry, and hence a survey of the magnetic geometries of ion-scale tangential discontinuities is necessary to determine whether these energetic particle populations contribute significantly to the solar wind heating as a whole.
The effect of the drift-kink instability to delay the growth of the tearing instability in the anti-parallel case suggests that two-dimensional studies of reconnection over-estimate the contribution of the tearing instability to the energisation of particles. Reconnection may therefore be unable to contribute a significant suprathermal population to the solar wind
We must also consider that the thin current sheets we simulate are initialised with idealised geometry in equilibrium, rather than arising from the evolution of turbulent fluctuations as is expected in the solar wind. Processes leading to the generation of thin current sheets in a turbulent medium may also contribute to particle energisation, and futher alter the ion velocity distribution from our ideal case.

\section{Observation}

Observational signatures of the three-dimensional evolution described in the preceding sections can be split between the two dominant instabilities described here: the tearing and drift-kink. As with the two-dimensional cases, the tearing instability can be identified by signatures of reconnection, including high temperature regions and beams in exhausts separated from the initial reconnection sites \citep{gosling2005}.

\begin{figure}
\includegraphics[width=\linewidth]{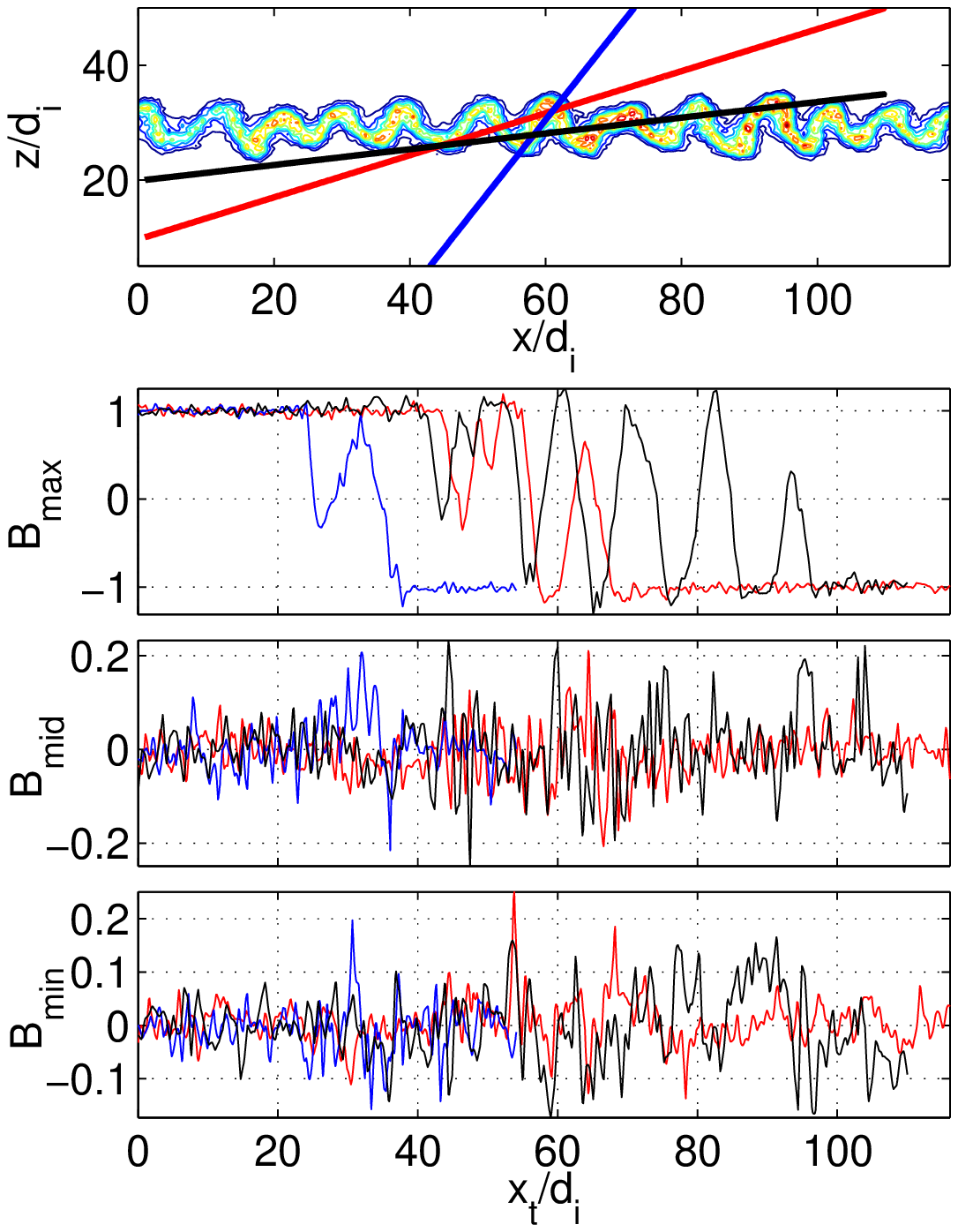}
\caption{\label{fig:obs_dk_single} Simulated spacecraft data for shallow crossings, showing minimum, medium and maximum variance components of the magnetic field. The chosen trajectories, with varying crossing angle, are shown in over contours of current sheet ion density in the top panel. Multiple crossings of the kinked current sheet are visible in the maximum variance component (equivalent to $B_y$).}
\end{figure}

\begin{figure}
\includegraphics[width=\linewidth]{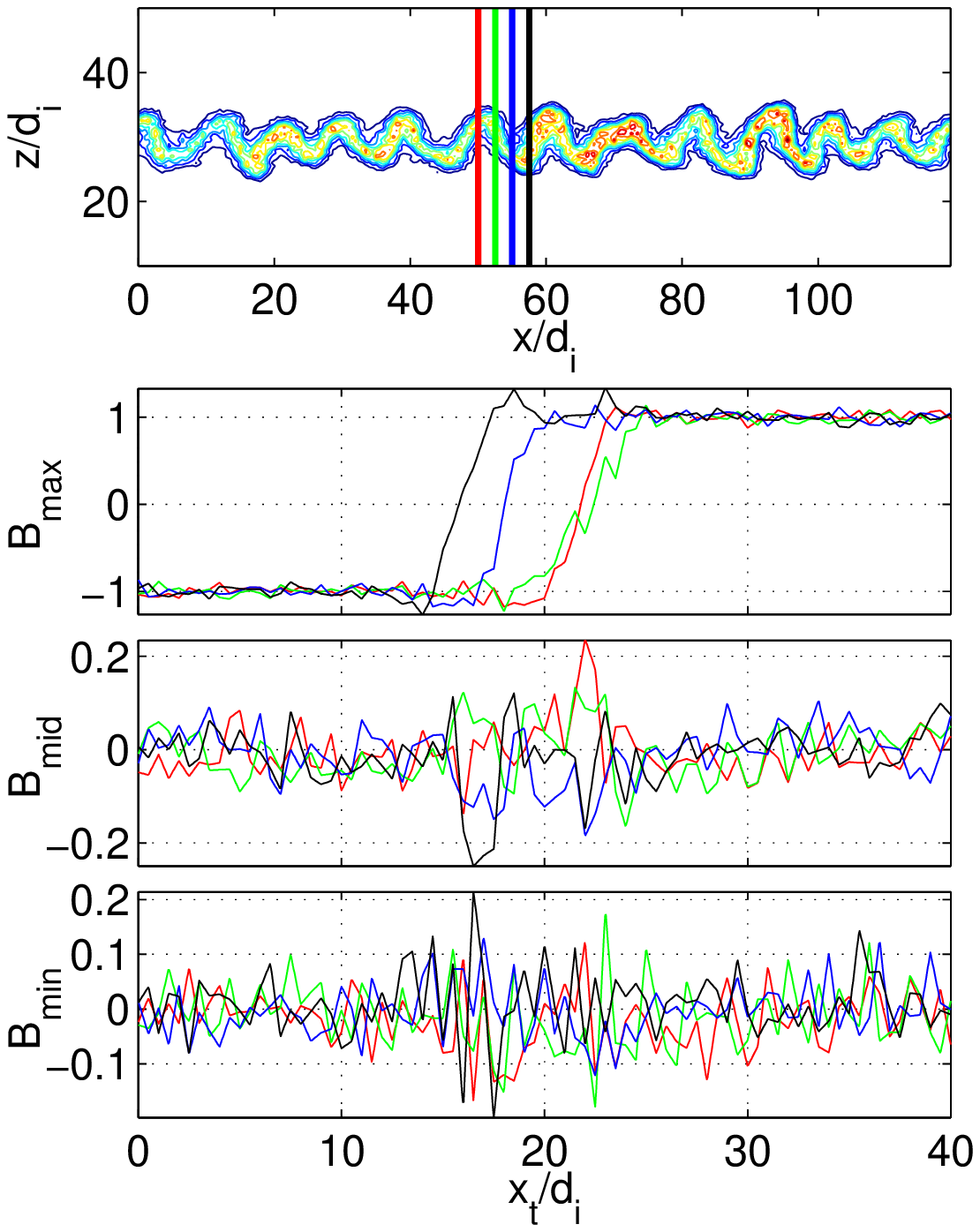}
\caption{\label{fig:obs_dk_multi} Simulated spacecraft data for perpendicular crossings, showing minimum, medium and maximum variance components of the magnetic field. The chosen trajectories, with varying crossing position, are shown over contours of current sheet ion density in the top panel. A set of multiple spacecraft crossings with variation in the position of the field rotation seen in the maximum variance component may be used to reconstruct a portion of a kinked current sheet.}
\end{figure}

For a single spacecraft, the drift-kink instability of thin current sheets can be observed for shallow crossings satisfying $ \phi > \tan^{-1} ( \lambda / 2 \pi A )$, where $\phi$ is the angle of the spacecraft trajectory from the current sheet normal, and $A, \lambda$ are the drift-kink amplitude and wavelength respectively. In this case, a single spacecraft will cross the kinked current sheet at least twice, as shown for a simulated crossing in Figure \ref{fig:obs_dk_single}, and the instability is therefore visible in magnetic field data. For multiple spacecraft systems such as CLUSTER, the drift-kink may also be observed as a periodic lag in expected crossing times observed between each spacecraft, as simulated in Figure \ref{fig:obs_dk_multi}. In this case, there is no condition on the approach angle, though some crossings may not be able to establish that the current sheet is more than singly curved, and there may be degenerate solutions for the wavelength of the kink. 

We stress that the simulations presented here represent ideal geometries. In reality, interaction with other solar wind structures, global effects such as curvature of the magnetic discontinuity over longer scales, and asymmetric expansion of the discontinuity with the solar wind are likely to further complicate the magnetic structure of kinked current sheet crossings. In addition, the internal structure of current sheets, such as magnetic islands formed by reconnection, can lead to a failure of minimum variance analysis to properly reconstruct the current sheet normal from signal data \citep{teh2011}.

\section{Conclusions}

The results presented in this paper have, in the first instance, been able to confirm the dependence on temperature anisotropy of the reconnection rate of thin current sheets as described in \citealt{matteini2013}: for higher perpendicular temperature, correlation of perturbations growing as a result of the ion cyclotron instability forces collapse of the current sheet and raises reconnection rates; for higher parallel temperature, anti-correlated perturbations as a result of the firehose instability lead to a kink and widening of the current sheet, reducing reconnection rates. We have shown that this relationship also holds for three-dimensional current sheets, however significant differences from the 2-D case arise. The most significant evolutionary feature introduced with the extension to three dimensions is the fast kink of the current sheets in the current carrying direction, characteristic of the growth of the drift-kink instability, a stream-type instability driven by the relative drift of the current carrying and background ion populations \citep{daughton1999}.

We have shown that the growth of the drift-kink instability suppresses the tearing instability by an effective widening of the current sheet after saturation. Importantly, this suggests that the thinnest ion-scale current sheets formed by turbulent fluctuations in the solar wind are transient, and therefore unable to support a significant, continuous rate of heating by reconnection. Hence, thinner current sheets are proportionately less able to heat the plasma via reconnection than wider structures, despite the increased growth rate of the tearing instability with decreasing current sheet thickness in the purely two-dimensional case. This effect is further compounded by the dependence of the growth rate of the drift-kink instability on the guide field angle: the current sheets most susceptible to tearing in two-dimensions, i.e. the anti-parallel case, have the greatest reduction in reconnection rate due to sheet widening by the drift-kink. In combination, these effects suggest that two-dimensional studies of reconnection in a turbulent solar wind are likely to over-estimate particle heating rates.

Although the results presented in this paper demonstrate that three-dimensional evolution of ion-scale current sheets is likely to suppress heating due to reconnection compared to the 2-D case, some important future extensions to this work remain before the net effect of the population of thin tangential discontinuities can be determined. Although we have accounted for interaction between a background and current sheet ion population, we have not accounted for the interaction of multiple current sheet populations with each other. This interaction may act as an additional source of turbulence in the solar wind, and is especially relevant to the plasma environment at the outer heliosphere, where pile up of plasma is expected to a produce a sectored heliosheath of parallel current sheets \citep{swisdak2013}. In particular, the limitation of the growth of magnetic islands in three-dimensions suggests that an extension of existing two-dimensional models of the heliopause will be necessary to recreate observations of the region by the Voyager spacecraft. Additionally, we have not yet discussed how the growth rate of tearing and drift-kink, and the interaction between the two instabilities, may vary with heliospheric distance and associated changes in plasma beta. These questions form the basis of our continuing research into this topic.

\acknowledgments

This work was supported by the UK Science and Technology Facilities Council grant ST/J001546/1.
The research leading to the presented results has received funding from the European Commission's
Seventh Framework Programme FP7 under the grant agreement SHOCK (project number 284515).
LM was funded by STFC grant ST/K001051/1.
This work used the DiRAC Complexity system, operated by the University of Leicester IT Services, which forms part of the STFC DiRAC HPC Facility (www.dirac.ac.uk). This equipment is funded by BIS National E-Infrastructure capital grant ST/K000373/1 and  STFC DiRAC Operations grant ST/K0003259/1. DiRAC is part of the National E-Infrastructure.

\end{document}